\documentclass[twocolumn]{aastex63}

\newcommand{\result}[1]{#1}

\graphicspath{{./}{fig/}}

\shorttitle{VLBI of L Dwarf Binary 2M J0746+2000AB}
\shortauthors{Zhang et al.}
\journalinfo{The Astrophysical Journal (preprint; accepted 2020 May 6)}

\begin{document}

\title{Multi-Epoch VLBI of L Dwarf Binary 2MASS J0746+2000AB: Precise Mass Measurements and Confirmation of Radio Emission from Both Components}

\author[0000-0002-6702-191X]{Qicheng Zhang}
\affiliation{Division of Geological and Planetary Sciences, California Institute of Technology, Pasadena, CA 91125, USA}

\author{Gregg Hallinan}
\affiliation{Department of Astronomy, California Institute of Technology, Pasadena, CA 91125, USA}

\author{Walter Brisken}
\affiliation{National Radio Astronomy Observatory, Socorro, NM 87801, USA}

\author{Stephen Bourke}
\affiliation{Department of Space, Earth and Environment, Chalmers University of Technology, Onsala Space Observatory, S-439 92 Onsala, Sweden}
\affiliation{Department of Astronomy, California Institute of Technology, Pasadena, CA 91125, USA}

\author{Aaron Golden}
\affiliation{School of Mathematics, Statistics \& Applied Mathematics, National University of Ireland Galway, Galway, Ireland}


\correspondingauthor{Qicheng Zhang}
\email{qicheng@cometary.org}

\begin{abstract}
Surveys have shown that up to one tenth of all ultracool dwarfs (UCDs) are appreciable radio emitters, with their emission attributed to a combination of gyrosynchrotron radiation and the electron cyclotron maser instability (ECMI). 2M J0746+2000AB is a close stellar binary comprised of an L0 and L1.5 dwarf that was previously identified as a source of 5~GHz radio emission. We used very-long-baseline interferometry (VLBI) to precisely track the radio emission over seven epochs in 2010--2017, and found both components to be radio emitters---the first such system identified---with the secondary component as the dominant source of emission in all epochs. The previously identified 2.07~h periodic bursts were confirmed to originate from the secondary component, although an isolated burst was also identified from the primary component. We additionally fitted the VLBI absolute astrometric positions jointly with existing relative orbital astrometry derived from optical/IR observations with Markov-chain Monte Carlo (MCMC) methods to determine the orbital parameters of the two components. We found the masses of the primary and secondary optical components to be \result{$0.0795\pm0.0003$~$M_\odot$} and \result{$0.0756\pm0.0003$~$M_\odot$}, respectively, representing the most precise mass estimates of any UCDs to date. Finally, we place a $3\sigma$ upper limit of 0.9~$M_{\text{jup}}$~au on the mass and separation of planets orbiting either of the two components.
\end{abstract}

\keywords{astrometry --- binaries: close --- stars: activity --- stars: low-mass --- techniques: high angular resolution --- techniques: interferometric}

\section{Introduction}

At the bottom of the stellar mass distribution sit the ultracool dwarfs (UCDs), a class of stellar and substellar objects typically defined by a spectral type of M7 or later \citep{kirkpatrick1997}. Their evolutionary pathways are largely set by their initial mass, with those above a minimum stellar mass threshold of $\sim$0.07~$M_\odot$ entering the main sequence as stable hydrogen-burning stars, and those below the threshold becoming deuterium-burning brown dwarfs, which more rapidly fade and cool after radiating away their formation energy \citep[e.g.,][]{kumar1963,hayashi1963,saumon2008,baraffe2015,dupuy2017}.

Evolutionary models have been developed to describe both low mass stars and brown dwarfs, and can be tested and refined through observation of nearby objects \citep[e.g.,][]{burrows1997,chabrier2000,saumon2008}. A near degeneracy in the the evolutionary tracks of UCDs hinders constraints on their mass or age from their temperature and luminosity, as is typically possible with more massive stars. Mass-luminosity-metallicity relationships have been developed for UCDs \citep[e.g.,][]{mann2019}, but are calibrated by a limited number of sources with well-constrained masses, which, unlike luminosity and metallicity, cannot be readily measured for isolated objects.

Close binary systems are well-suited as model calibrators due to their measurable orbits which provide their dynamical masses, and their presumed co-evolution which eliminates age as a confounding factor between the components. 2M J0746+2000 is a nearby spectral class L source cataloged by the Two Micron All Sky Survey \citep[2MASS;][]{skrutskie2006} that was resolved to be a close binary system comprised of an L0 primary component (2M J0746+2000A) and an L1.5 secondary component (2M J0746+2000B) by a Hubble Space Telescope (\textit{HST}) survey of L dwarfs \citep{reid2001}. \citet{bouy2004} provided additional relative astrometric observations, extending the observational baseline to four years, and found, through model fitting of the relative orbit, a system mass of $0.146_{-0.006}^{+0.016}$~$M_\odot$ ($2\sigma$ bounds)---the first dynamical mass measurement of an L dwarf system. \citet{konopacky2010} further extended the baseline to eight years with relative astrometry from \textit{HST} and ground-based adaptive optics imagery, and found a system mass of $0.151\pm0.003$~$M_\odot$.

The actual individual masses of the components, however, require the absolute orbits in an inertial frame, which can be established through radial velocity measurements or absolute astrometry. \citet{konopacky2010} performed radial velocity observations of the two components, but were prevented from meaningfully constraining the relative masses by the low precision of the measured velocities, especially as they sampled a portion of the orbit with low relative radial velocities. Absolute astrometry of resolved optical/IR imagery, meanwhile, is hindered by the limited availability of fixed, extragalactic reference sources ideally used to anchor astrometry to an inertial frame, due the relatively restrictive field of view provided by adaptive optics, and to a lesser extent, \textit{HST}. Calibration to faint field stars can introduce significant systematic errors, particularly if only a small number are available, although the recent availability of high precision field star astrometry \citep{gaia2018} substantially mitigates these errors when at least a few such field stars are available, as is typically true for \textit{HST} imagery \citep[e.g.,][]{bedin2018}.

Adaptive optics, on the other hand, may not provide any calibrated field stars. As one workaround, \citet{cardoso2012} used point spread function (PSF) fitting of unresolved wide field images with large numbers of well-calibrated field stars, informed by higher precision relative astrometry provided by adaptive optics, to precisely constrain the absolute orbit and individual masses of the T dwarf binary $\varepsilon$~Indi~B, which were were later refined by similar methods \citep[e.g.,][]{dieterich2018}. \citet{harris2015} and \citet{dupuy2017} later applied an equivalent method to 2M J0746+2000AB, fitting the absolute motion of the system's photocenter in wide field imagery jointly with higher precision relative astrometry. The latter paper found individual masses of $0.0787_{-0.0014}^{+0.0013}$~$M_\odot$ and $0.0749\pm0.0013$~$M_\odot$ for 2M J0746+2000A and B, respectively---far superior to the precision possible with radial velocity measurements. Uncertainties were now constrained at this point by the relative optical variability of the components, systematic biases in proper motion and parallax of the field stars, and the comparatively poor resolution of the wide field imagery that requires a large number of observations to overcome.

By happenstance, the 2M J0746+2000 system is also one of a small fraction \citep[$\sim$10\%;][]{route2016,pineda2017} of ultracool dwarfs detected in radio surveys to date \citep{antonova2008,berger2009}. Radio emission has been detected from objects extending down to spectral type $\sim$T6.5 \citep{route2012}, including from an object with a possible mass of just $\sim$13~$M_{\text{Jup}}$ \citep{gagne2017,kao2018}. All members of the radio-detected sample have been found to be sources of both quiescent radio emission, typically attributed to gyrosynchrotron emission \citep{berger2002}, and sources of periodic pulsed radio emission, consistent with that expected from electron cyclotron maser instability \citep[ECMI;][]{hallinan2007,hallinan2008}. The latter is a likely manifestation of auroral currents in a large-scale magnetosphere, and can be accompanied by optical and Balmer line periodic variability \citep{hallinan2015}. The exact electrodynamic engine powering auroral activity in a small fraction of ultracool dwarfs, all of which are rapid rotators, remains uncertain. Possibilities include an orbiting exoplanet \citep{hallinan2015, pineda2017}, analogous to the role played by Io in a component of the Jovian decametric emission \citep{zarka1998} or a breakdown of co-rotation of plasma in the middle magnetosphere, analogous to the role Iogenic plasma plays in generating the main Jovian aurora oval \citep{schrijver2009,nichols2012}.

2M J0746+2000 presents a particularly curious case. It was detected initially as a radio source by \citet{antonova2008} and subsequently confirmed to be a pulsing source by \citet{berger2009}. \citet{berger2009} also observed periodic H$\alpha$ emission from the combined system with the same 2.07~h period seen in the radio. Using spatially unresolved archival $v\sin i$ measurements of the binary, and assuming the rotation axis to be orthogonal to the orbital plane, \citet{berger2009} attributed the 2.07~h period to 2M J0746+2000A's rotation and argued that its radius must therefore be $\sim$30\% smaller than expected from theoretical models. However, individual $v\sin i$ measurements of each component \citep{konopacky2012} together with the 3.3~h period measured from spatially unresolved broadband optical photometry---attributed to the rotation period of the non-radio component---later established that 2M J0746+2000B is likely the dominant radio-emitting source, and that the radii of both components are therefore consistent with theoretical expectations \citep{harding2013b}. The fact that the secondary appears to produce both radio emission and H$\alpha$ emission, as implied by the shared 2.07~h period, while the primary is dominates the broadband optical variability, is unexpected. \citet{harding2013a} showed that most radio-detected ultracool dwarfs produce periodic optical variability---attributed to auroral activity---similar to that responsible for periodic pulsing radio emission and periodic H$\alpha$ emission \citep{hallinan2015}. However, the non-radio emitting component in the 2M J0746+2000 system was the one detected as the dominant source of optical variability. One possibility is that the optical variability of the non-radio emitting component is due to inhomogeneities in its atmosphere, as has been observed for a sample of L and T dwarfs \citep{bailer-jones2001,gelino2002,maiti2005,littlefair2008,harding2013b}. \citet{harding2013a} suggested that the non-radio emitting component (likely 2M J0746+2000A) may also be a radio emitter, albeit at lower flux densities than detectable to date.

Importantly, the radio emission enables the use of very-long-baseline interferometry (VLBI) to precisely pinpoint the emission source relative to extragalactic references sources that are, for all practical purposes, fixed in inertial reference frames. VLBI can isolate the emission to individual components of the binary, and trace their absolute motion in the sky with extremely high precision. \citet{dupuy2016} previously employed such a method to obtain absolute astrometry, and subsequently constrain the orbits and individual masses of the pre-main sequence LSPM J1314+1320AB system whose secondary component was found to be radio-emitting.

2M J0746+2000AB presents an opportunity to similarly investigate an older system with components much closer to the minimum stellar mass threshold. In this manuscript, we present 5~GHz VLBI observations of 2M J0746+2000AB from seven epochs spanning seven years. We locate and discuss the properties of sources of emission matching the expected motion of the two components. We then jointly fit their positions together with previously published relative astrometry reduced by \citet{dupuy2017} and \citet{bouy2004} to tightly constrain the absolute motion of 2M J0746+2000A and B and determine their individual dynamical masses. We discuss our observations in the context of earlier observations, and establish limits on the presence of planets in the system set by the astrometric residuals.

\section{Observations}

\begin{deluxetable*}{cccc}
\centering
\tablecaption{Epochs considered in this analysis, and the stations contributing to each epoch}

\tablehead{\colhead{Epoch} & \colhead{Time (UT)} & \colhead{Time On Target (h)} & \colhead{Antennas\tablenotemark{a}}}
\startdata
GH009A & 2010 Mar 21 21:12--22 01:58 & 2.73 & VLBA--HN--MK+EF+MC+NT+WB+GB\\
BH181A & 2013 Feb 5 01:35--06:24 & 3.12 & VLBA--SC\\
BH181B & 2013 Sep 20 10:50--15:41 & 3.27 & VLBA\\
BH181C & 2015 Aug 28 12:23--17:09 & 3.44 & VLBA--SC\\
BH181D & 2016 Feb 23 00:08--04:56 & 3.23 & VLBA\\
BH181E & 2016 Aug 28 12:08--16:53 & 3.38 & VLBA--PT--SC\\
BH181F & 2017 Feb 21 00:08--04:54 & 3.29 & VLBA\\
\enddata
\tablenotetext{a}{Station codes (``--'' prefixes the codes of excluded VLBA stations and ``+'' prefixes stations not part of the 10-station VLBA): SC=St. Croix, HN=Hancock, MK=Mauna Kea (VLBA); EF=Effelsberg, MC=Medicina, NT=Noto, WB=Westerbork (EVN); GB=Green Bank}
\label{tab:obslist}
\end{deluxetable*}

\begin{figure*}
\centering
\includegraphics[width=\linewidth]{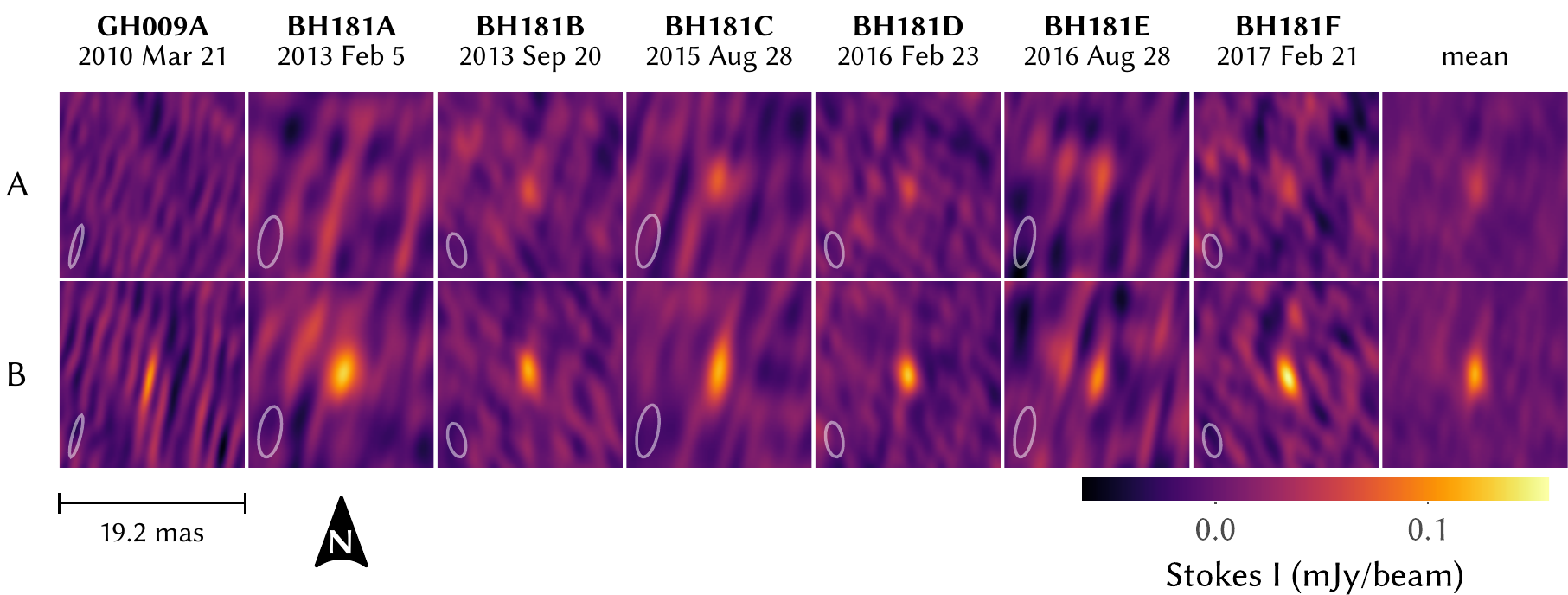}
\caption{Images of 5~GHz Stokes $I$ emission from 2M J0746+2000A (top row) and B (bottom row), and the mean stacks over all epochs (right). Ellipses indicate the shape and size of the beam at each epoch.}
\label{fig:src_ab}
\end{figure*}

Radio observations were conducted of 2M J0746+2000AB with the Very Long Baseline Array (VLBA) and the Green Bank Telescope (GBT) over seven epochs in 2010--2017, as summarized in Table~\ref{tab:obslist}. The first epoch (labeled ``GH009A'') also incorporates concurrent observations by stations in the European VLBI Network (EVN). The data quality for this particular epoch was poor, and the corresponding astrometry was downweighted by a procedure elaborated in section \ref{sec:orbfit}. Due to a recently uncovered long-term problem with phase coherence for GBT, all visibilities involving the dish were flagged for all epochs, significantly reducing the sensitivity relative to original projections. The sensitivity was still sufficient for the primary goal of dynamical mass measurement and the investigation of radio emission from both components. The loss in sensitivity, however, did hinder measurement of the relative position of left and right circularly polarized pulses of radio emission within the magnetosphere of 2M J0746+2000B and the investigation of planetary bodies within the binary system. 

Each epoch spans a 5~h period and covers a series of rapid-switching phase-referencing observations alternating between 4~min on the 2M J0746+2000 field and 1~min on the reference source, J0750+1823, $\sim$2$\arcdeg$ away. Eight C band spectral channels were used, covering the frequency range 4.85--5.11~GHz in dual polarization.

Additionally, a compact extragalactic radio source was identified north of 2M J0746+2000AB at International Celestial Reference System (ICRS) R.A. $07^{\text{h}}46^{\text{m}}42^{\text{s}}\llap{.}982532\pm0^{\text{s}}\llap{.}000006$ and decl. $+20\arcdeg00'37''\llap{.}982532\pm0''\llap{.}00009$ within the primary beam when targeting 2M J0746+2000. A secondary phase center was requested to correlate the in-beam source in the BH181A--F epochs to correct phase offsets associated with the calibration. However, due to correlator technical issues beyond our control, this data was only collected for BH181B, BH181E, and BH181F, so the in-beam source is unavailable for the other four epochs.

The raw visibility data was initially calibrated with the Astronomical Image Processing System (\textsc{aips}) software package \citep{greisen1990}. The calibrated visibilities were subsequently imaged and deconvolved with the \texttt{clean} utility in the Common Astronomy Software Applications (\textsc{casa}) package \citep{jaeger2008}.

\section{Radio Emission Properties}

\begin{deluxetable*}{cccccccc}
\tablecaption{ICRF R.A. ($\alpha$) and decl. ($\delta$) of the two radio sources, their uncertainties, the covariance of the uncertainties, and the mean Stokes $I$ flux (or $3\sigma$ bound, for non-detections) at every epoch\label{tab:posfit}}

\tablehead{\colhead{Epoch} & \colhead{Component} & \colhead{$\alpha$ (07:46:XX)} & \colhead{$\sigma_\alpha\cos\delta$ (mas)} & \colhead{$\delta$ (+20:00:XX)} & \colhead{$\sigma_\delta$ (mas)} & \colhead{cov (mas$^2$)} & Flux (mJy)}
\startdata
GH009A\tablenotemark{a}\tablenotemark{b} & A & \nodata & \nodata & \nodata & \nodata & \nodata & $<$0.042\\
& B & 42.2276733 & 0.160 & 31.330407 & 0.382 & $-0.0280$ & $0.119\pm0.026$\\
BH181A\tablenotemark{b} & A & \nodata & \nodata & \nodata & \nodata & \nodata & $<$0.053\\
& B & 42.1629606 & 0.167 & 31.219139 & 0.381 & $-0.0195$ & $0.161\pm0.027$\\
BH181B\tablenotemark{c} & A & 42.1491173 & 0.154 & 31.347589 & 0.354 & $+0.0293$ & $0.078\pm0.017$\\
& B & 42.1553189 & 0.121 & 31.200961 & 0.241 & $+0.0118$ & $0.131\pm0.017$\\
BH181C\tablenotemark{b} & A & 42.0988160 & 0.202 & 31.120543 & 0.591 & $-0.0583$ & $0.077\pm0.020$\\
& B & 42.1025746 & 0.158 & 31.231021 & 0.328 & $-0.0156$ & $0.127\pm0.017$\\
BH181D\tablenotemark{b} & A & 42.0813946 & 0.183 & 31.104002 & 0.447 & $+0.0247$ & $0.059\pm0.016$\\
& B & 42.0801643 & 0.149 & 31.224959 & 0.244 & $+0.0055$ & $0.124\pm0.016$\\
BH181E\tablenotemark{c} & A & 42.0777828 & 0.167 & 31.073825 & 0.628 & $-0.0666$ & $0.076\pm0.022$\tablenotemark{d}\\
& B & 42.0715563 & 0.174 & 31.167184 & 0.667 & $-0.0753$ & $0.084\pm0.026$\tablenotemark{d}\\
BH181F\tablenotemark{c} & A & 42.0599523 & 0.171 & 31.082929 & 0.462 & $+0.0368$ & $0.071\pm0.022$\\
& B & 42.0499620 & 0.109 & 31.135131 & 0.219 & $+0.0069$ & $0.166\pm0.021$\\
\enddata
\tablenotetext{a}{Epoch downweighted in orbital fit as potential outlier due to suspect calibration}
\tablenotetext{b}{Positional uncertainties include a symmetric \result{0.19~mas} (0.13~mas in $\alpha\cos\delta$ and $\delta$) phase calibration uncertainty}
\tablenotetext{c}{Positions corrected with in-beam reference source, but stated uncertainties include a symmetric \result{0.13~mas} (0.09~mas in $\alpha\cos\delta$ and $\delta$) systematic uncertainty in in-beam source position}
\tablenotetext{d}{May be underestimated due to suspect calibration in several bands distributing flux into one or more secondary peaks.}
\end{deluxetable*}

\begin{figure}
\centering
\includegraphics[width=\linewidth]{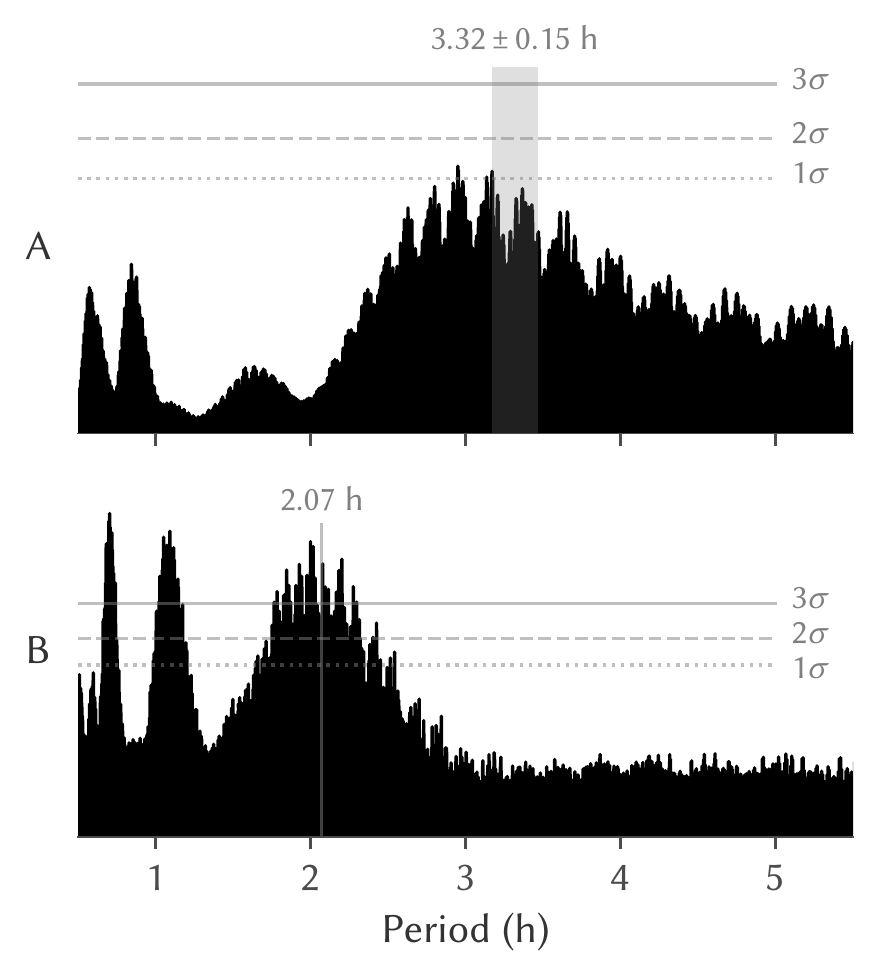}
\caption{Lomb--Scargle periodograms of the 5~GHz Stokes $I$ flux from the A (top) and B (bottom) sources over the epochs where emission was detected from each component. Horizontal lines mark the $1\sigma$, $2\sigma$, and $3\sigma$ false alarm probability levels estimated by the method of \citet{baluev2008}, while vertical bars mark the rotation periods previously measured by \citet{harding2013b} and \citet{berger2009} for A and B, respectively. Note the presence of higher order harmonics, indicative of non-sinusoidal flux variation, as expected given the presence of bursts.}
\label{fig:period}
\end{figure}

\begin{figure*}
\centering
\includegraphics[width=0.75\linewidth]{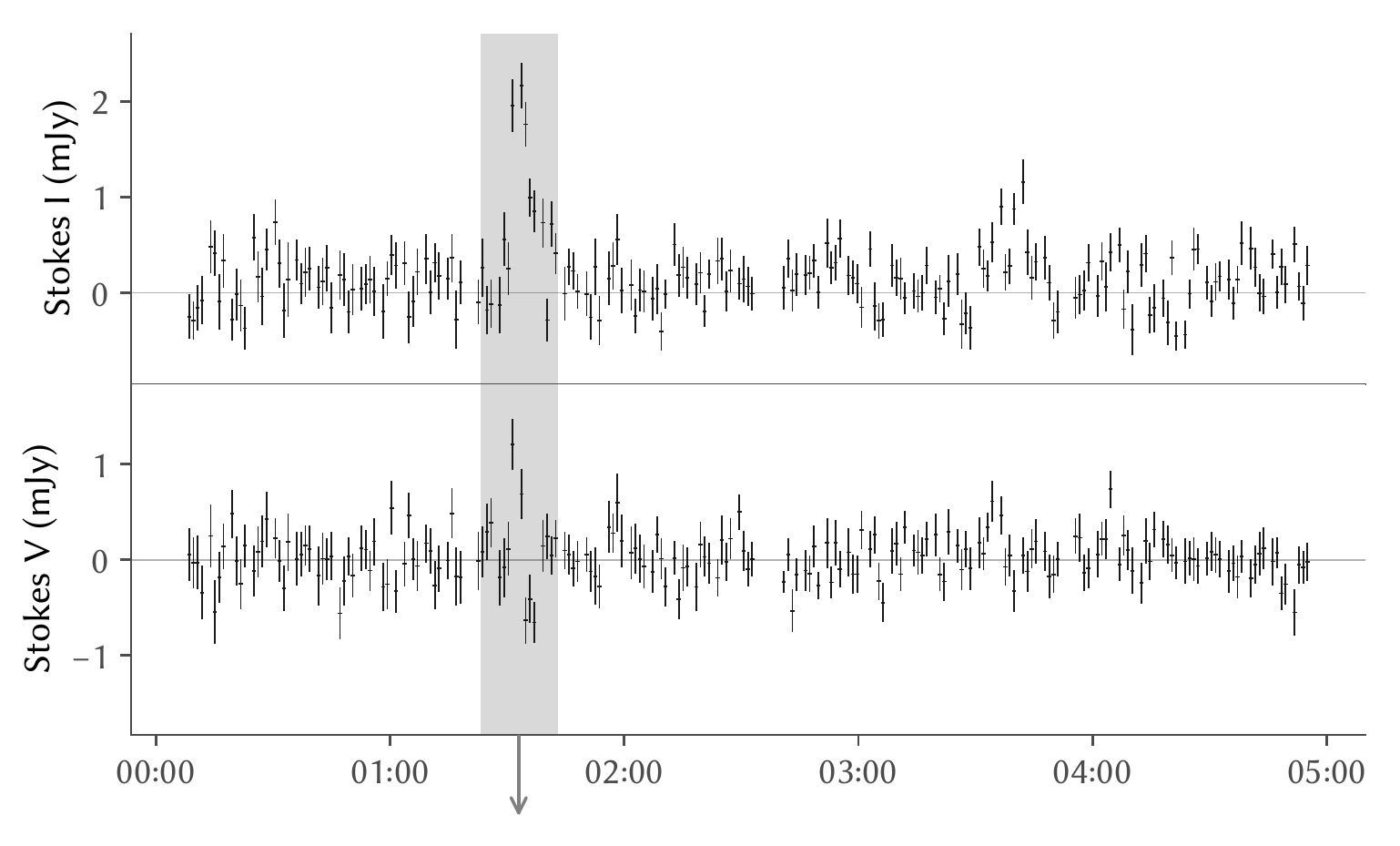}
\includegraphics[width=0.75\linewidth]{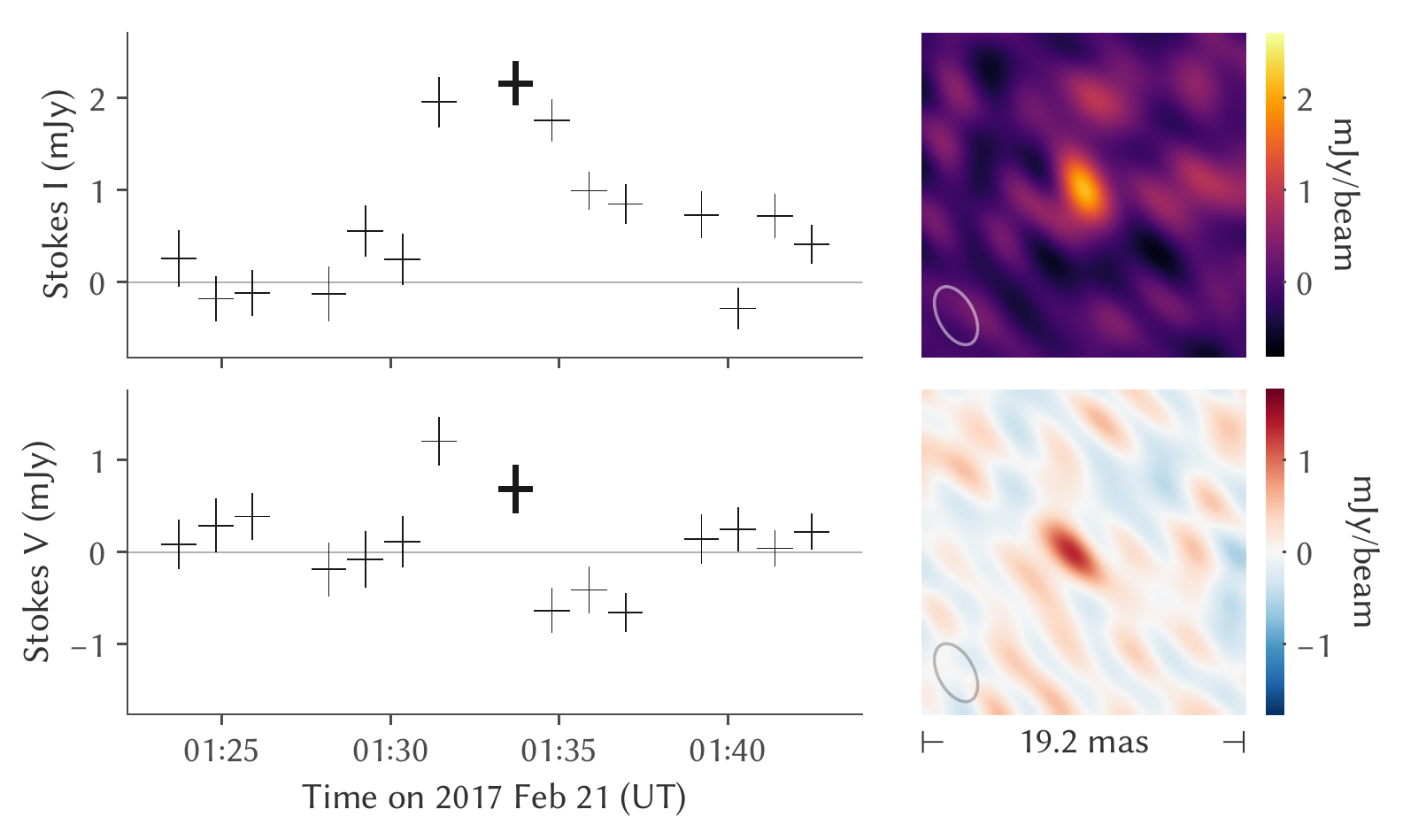} 
\caption{Stokes $I$ and $V$ light curves of 2M J0746+2000B in the BH181F epoch showing two bursts at 01:34 and 03:38 (top), with an inset of the 01:34 burst (shaded region) imaged at the bolded point (bottom). An animation of the burst inset that cycles through images corresponding to all points on the inset light curve is provided as the ancillary file \texttt{BH181FB\_burst.webm}.}
\label{fig:BH181FB_burst}
\end{figure*}

\begin{figure}
\centering
\includegraphics[width=\linewidth]{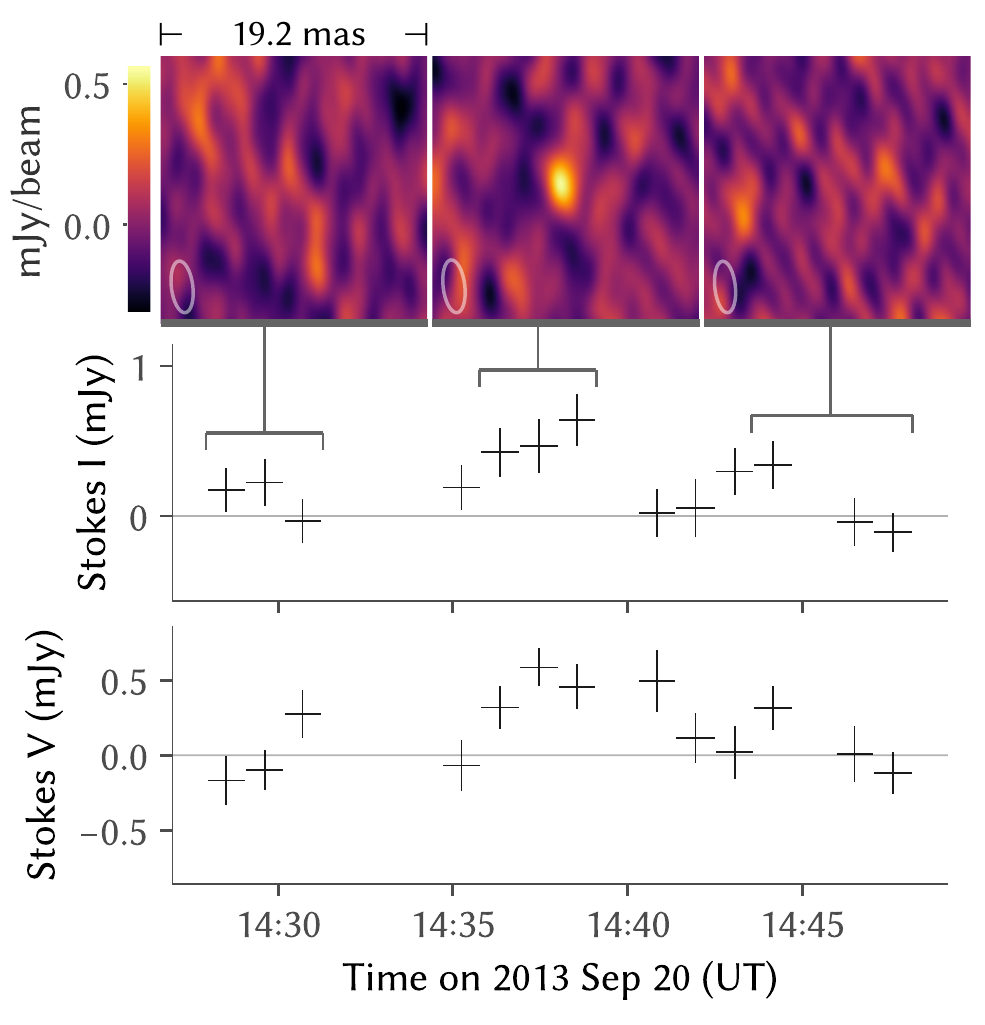}
\caption{Stokes $I$ and $V$ light curves of a polarized burst from 2M J0746+2000A in the BH181A epoch, with Stokes $I$ images from before, during, and after the burst corresponding to the indicated points.}
\label{fig:BH181BA_burst}
\end{figure}

Only a single clear source was initially identified near the predicted optical/IR position of the 2M J0746+2000 system in images of each of the epochs. The motion of this source is consistent with the optical/IR motion of the B component providing the first direction confirmation of this component as the main contributor of radio emission from the binary, and confirming the conclusions reached by \citet{konopacky2012} and \citet{harding2013b}.

The corresponding position of the A component at each epoch was then identified using the previously published relative orbits. Figure~\ref{fig:src_ab} presents images centered on the expected A component positions and corresponding B component sources, showing a weak but clear signal at the expected A component position in all but the first two epochs. Neither source is clearly distinguishable from a point source in any of the images.

Lomb--Scargle periodograms \citep{lomb1976,scargle1982} were used to analyze periodicity in the radio emission of both radio sources. Figure~\ref{fig:period} shows periodograms computed from the flux at 2~s intervals at the best fit positions across all epochs where each source was detected. The $1\sigma$, $2\sigma$, and $3\sigma$ false alarm probability levels, conservatively estimated by the method of \citet{baluev2008}, are included for reference, and represent the likelihood the maximum power exceeds these levels if the data series were normally distributed.

The rotation period of the A component was previously attributed to a rotation period of $3.32\pm0.15$~h by \citet{harding2013b}, therefore leading to the conclusion that the $2.072\pm0.002$~h radio bursts found by \citet{berger2009} originate from the B component. The periodograms are consistent with this conclusion, and show a $>$3$\sigma$ peak near a 2~h period from the B source and a weak peak at 3~h from the A source at just over the $1\sigma$ false alarm probability level. Additionally, both sources show higher order harmonics at reciprocal integer fractions of their respective periods, reflecting the non-sinusoidal variation in the radio emission, such as from bursts.

As the signal-to-noise ratio (SNR) of the period data is much lower than that of \citet{harding2013b} and \citet{berger2009}, we cannot further constrain the rotation periods of the components to better precision, and so do not attempt to rigorously model the periodicity. However, we can very crudely estimate the periods and their uncertainties as the simple mean and standard deviation of the period derived from the three strongest harmonics of each source (first, fourth, and fifth for A; first, second and third for B), and find $3.2\pm0.2$~h and $2.1\pm0.1$~h for the A and B sources, respectively, which is consistent with the published values.

In contrast to the observations of \citet{berger2009}, no bursts exceeding 10~mJy were detected in any of the seven epochs. This high degree of variability of the luminosity of the periodic pulsed emission has been previously reported for radio detected ultracool dwarfs \citep{hallinan2007}. Weaker $\sim$1~mJy bursts, however, were observed from the B source in all epochs, and generally recurred with a similar 2.1~h period. Figure~\ref{fig:BH181FB_burst} shows two such bursts in the light curve of the B source over the BH181F epoch, with the first being the strongest burst observed across all epochs. Unlike the 100\% circularly polarized bursts observed by \citet{berger2009}, this burst appears to be unpolarized for at least part of the burst. This effect can be explained by two overlapping bursts of opposite polarization, with the left polarized (Stokes $V<0$) component lagging the stronger right polarized ($V>0$) component by several minutes. Similar behavior was previously observed from this system by \citet{lynch2015}. 

Additionally, a single 0.5~mJy right polarized burst was observed from the A source in the BH181B epoch. A light curve of this event, together with images created at the indicated times, are presented in Figure~\ref{fig:BH181BA_burst}. Similar images made from the data 3.0--3.5~h earlier in the same epoch---one A component rotation period earlier---revealed no clear signal, and no additional bursts were discerned in any of the other epochs.

\subsection{Implications of Radio Emission from Both Components}

2M J0746+2000AB is now only the second radio emitting UCD binary system investigated with VLBI, after the M7 binary LSPM J1314+1240AB \citep{dupuy2016}, and is the first identified to have two radio emitting components. Our observed C band luminosities for both 2M J0746+2000A and B is comparable to that observed from the other radio-emitting L dwarfs, and higher than that observed from T dwarfs \citep{pineda2017}, and the radio emission from both components would individually have been detected by surveys had they been spatially separated. Only $\sim$10\% of surveyed UCDs have been detected as radio emitters \citep{route2016}. If the presence or absence of radio emission were independent between binary components, the probability that at least one of the two observed systems would have two radio emitters would only be $\sim$20\%. While not statistically significant by typical thresholds, this result motivates speculation on the possible correlation of radio emission presence between the individual components of multiple UCD systems.

In assessing such a correlation, we note that in each of the two observed systems, the binary components are similar in mass and can be reasonably assumed to have formed at the same time. The absence of a sample of objects with known mass or age prevents a robust assessment of a correlation between these properties and the presence of radio emission. However, a correlation has been shown with $v\sin i$ \citep{mclean2012}, with evidence for a sharp rise in the detection fraction at $v\sin i>40$~km~s$^{-1}$ \citep{pineda2017}, corresponding to a 3.1~h rotation period for inclinations close to $90\arcdeg$. We note an additional bias towards detection for inclinations significantly larger than $0\arcdeg$ \citep{pineda2017}. We further note that both components of 2M J0746+2000 are rapid rotators, with rotation periods of $\sim$3.3~h and 2.07~h, for the A and B components, respectively, while the rotation axes of both components are aligned to within $10\arcdeg$ of the orbital pole, as consistent with many binary formation pathways \citep{harding2013b}.

The detection of both components is therefore not surprising, given the increased detection fraction for rapid rotators. Under the assumption of coevality and formation via fragmentation of a molecular cloud core, similar mass components in a tight binary are expected to have similar rotational velocities, with rotation axes approximately aligned with the rotation axis of the cloud core \citep{bate1997}. \citet{konopacky2012} showed that a number of UCD binaries, including 2M J0746+2000AB, have somewhat discrepant rotational velocities, with the secondary rotating somewhat faster than the primary. Nonetheless, there remains a strong correlation between the rotation rates of binary components. VLBI investigation of the components of the remaining confirmed radio emitting UCD binaries LP 349-25 \citep{phan-bao2007} and 2M J1315-2649 \citep{burgasser2012} is therefore warranted, with the optical light curve of the former suggesting possible radio emission contribution from both of its components \citep{harding2013b}.

Finally, we note that magnetic field interaction between the binary components is unlikely to be responsible for the observed radiation. A 5~GHz gyrofrequency, as observed with 2M J0746+2000AB, occurs in a magnetic field strength of $\sim$2~kG. We can then model the system as a pair of dipole magnetic fields, scaled from an optimistic $\sim$10~kG surface field at a radius of $\sim$1~$R_{\text{Jup}}$, with electrons supplied by an interstellar medium with electron density $\sim$1~cm$^{-3}$. Using the plasma interaction model of \citet{zarka2007}, the flux radiated by electrons accelerated by the pair's interacting dipolar magnetic fields would be $\sim$1~nJy on Earth---far below that of the observed signal and the detection threshold.

\section{Orbital Fitting}
\label{sec:orbfit}

\begin{deluxetable*}{cccccccc}
\tablecaption{Burst and ex-burst\tablenotemark{a} astrometry used for orbital fit in place of corresponding full epoch astrometry (Table~\ref{tab:posfit})\label{tab:bursts}}

\tablehead{\colhead{Time (UT)} & \colhead{Polarization} & \colhead{$\alpha$ (07:46:XX)} & \colhead{$\sigma_\alpha\cos\delta$ (mas)} & \colhead{$\delta$ (+20:00:XX)} & \colhead{$\sigma_\delta$ (mas)} & \colhead{cov (mas$^2$)} & Flux (mJy)}
\startdata
\noalign{\vspace{0.8ex}}
\multicolumn{8}{c}{A component burst in BH181B (Figure \ref{fig:BH181BA_burst})}\\
\noalign{\vspace{0.8ex}}
\hline
2013 Sep 20 14:36--14:39\phantom{\tablenotemark{a}} & RR & 42.1491356 & 0.153 & 31.347923 & 0.354 & $+0.0115$ & $0.96\pm0.19$ \\
\phantom{2013 Sep 20 }10:50--15:41\tablenotemark{a} & I & 42.1491155 & 0.157 & 31.347490 & 0.372 & $+0.0314$ & $0.071\pm0.012$ \\
\hline
\noalign{\vspace{0.8ex}}
\multicolumn{8}{c}{B component burst in BH181F (Figure \ref{fig:BH181FB_burst})} \\
\noalign{\vspace{0.8ex}}
\hline
2017 Feb 21 01:33--01:37\phantom{\tablenotemark{a}} & LL & 42.0499569 & 0.208 & 31.135461 & 0.274 & $+0.0292$ & $1.85\pm0.23$ \\
\phantom{2017 Feb 21 }01:31--01:35\phantom{\tablenotemark{a}} & RR & 42.0499620 & 0.195 & 31.134938 & 0.246 & $+0.0221$ & $2.46\pm0.27$ \\
\phantom{2017 Feb 21 }00:08--04:54\tablenotemark{a} & I & 42.0499637 & 0.115 & 31.135117 & 0.249 & $+0.0095$ & $0.128\pm0.014$ \\
\enddata
\tablenotetext{a}{ex-burst: full epoch, excluding data included in burst astrometry}
\end{deluxetable*}

The position of both radio sources were astrometrically fitted in the integrated image of each epoch in which they are detected. Fitting was performed with a 2D Gaussian function with dimensions and orientation matching that of the restoring beam determined by \textsc{casa}. Fit uncertainties were estimated using the correlated noise formulas \citet{condon1997} as applied by \textsc{casa}'s \texttt{imfit} utility.

The positions of the in-beam source was similarly determined in the three epochs for which it was available. Subtracting the fit uncertainty from the root mean square (rms) variation of the fitted positions in quadrature leaves an estimated phase calibration error of \result{$\sigma_{\text{cal}}=0.19$~mas}, treated as symmetric in R.A. and decl. This $\sigma_{\text{cal}}$ was then added in quadrature to the fit uncertainties of the target radio sources for the four epochs where the in-beam source is not available.

The position of the in-beam source is taken to be the mean of the three fitted positions, with an assumed symmetric uncertainty \result{0.13~mas} derived from the rms. The displacement of the in-beam source position from the mean at each of the three epochs---attributed to an offset error of the phase calibration at 2M J0746+2000---was added to the corresponding fitted target source position, with the in-beam position uncertainty added in quadrature. The in-beam uncertainty also introduces a covariance between in-beam corrected positions equal to the square of the in-beam uncertainty, 0.017~mas$^2$. This correlation does not appreciably affect the final solution as the covariance is dwarfed by the square of the R.A. and decl. uncertainties added in quadrature, so is excluded for computational simplicity. The final corrected positions and uncertainties of the components in all seven epochs with respect to the International Celestial Reference Frame (ICRF) are given in Table~\ref{tab:posfit}.

Additionally, the dominant A component burst in the BH181B epoch and B component burst in BH181F each peak at a signal-to-noise ratio (SNR) comparable to the SNR of the source integrated over the full epoch, so astrometric positions can be measured for each burst individually to a precision comparable with that measured for the full epoch. As the bursts are polarized, SNR and thus astrometric precision are further improved by separately fitting the RR and LL signal. A physical separation of two polarized components of each burst on the order of the stellar radii ($\sim$40~$\mu$as) may theoretically be present, reflecting the separation of emission from the two magnetic poles. We neglect this physical separation in our analysis as its scale is far below the resolution of our astrometry.

Finally, the bursts contribute a minute fraction of the total flux recorded for the respective sources and epochs. We can therefore create a copy of the full epoch data with the bursts removed and measure an additional ``ex-burst'' position, which will have a precision only slight worse than that obtained with the full epoch. Errors in the burst and ex-burst positions are correlated by the 0.13~mas uncertainty of the in-beam source position used to calibrate the astrometry of BH181B and BH181F, which is similar to the correlation between the corresponding full epoch astrometry and handled in an identical manner. The in-beam source corrected burst and ex-burst astrometry are given in Table~\ref{tab:bursts}, and used in place of the full epoch BH181B A component and BH181F B component positions Table~\ref{tab:posfit} for the orbit fitting procedure that follows.

\begin{deluxetable}{lcccc}
\tablecaption{Previously published optical/IR astrometry of 2M J0746+2000B relative to A used for orbital fit\label{tab:relastrom}}

\tablehead{\colhead{Time (UT)} & \colhead{Separation (mas)} & \colhead{Pos. Angle ($\arcdeg$)} & \colhead{Ref.}}
\startdata
2000 Apr 15.34 & $217.8\pm2.9$ & $168.8\pm0.5$ & (1)\\
2002 Feb 7.41 & $121\pm8$ & $\phantom{0}86\pm4$ & (2)\\
2002 Oct 21.97 & $121.78\pm0.10$ & $\phantom{0}33.80\pm0.28$ & (1)\\
2003 Mar 22.06 & $123.5\pm2.1$ & $\phantom{00}4.6\pm1.0$ & (2)\\
2003 Dec 4.64 & $126.5\pm1.8$ & $317.9\pm0.7$ & (2)\\
2004 Jan 9.79 & $134.5\pm3.0$ & $311.1\pm1.2$ & (2)\\
2007 Dec 1.62 & $334.13\pm0.19$ & $223.64\pm0.02$\tablenotemark{a} & (1)\\
2007 Dec 1.63 & $334.1\pm0.5$ & $223.59\pm0.06$\tablenotemark{a} & (1)\\
2008 Dec 18.48 & $351.09\pm0.29$ & $214.50\pm0.20$\tablenotemark{a} & (1)\\
2008 Dec 18.50 & $347.97\pm0.15$ & $205.95\pm0.02$\tablenotemark{a} & (1)\\
\enddata
\tablerefs{(1) \citet{dupuy2017}; (2) \citet{bouy2004}}
\tablenotetext{a}{Offset from published values by $-0\arcdeg\llap{.}50$, with a recently corrected implementation of the \citet{yelda2010} NIRC2 calibration \citep{bowler2018}.}
\end{deluxetable}

\begin{figure}
\centering
\includegraphics[width=\linewidth]{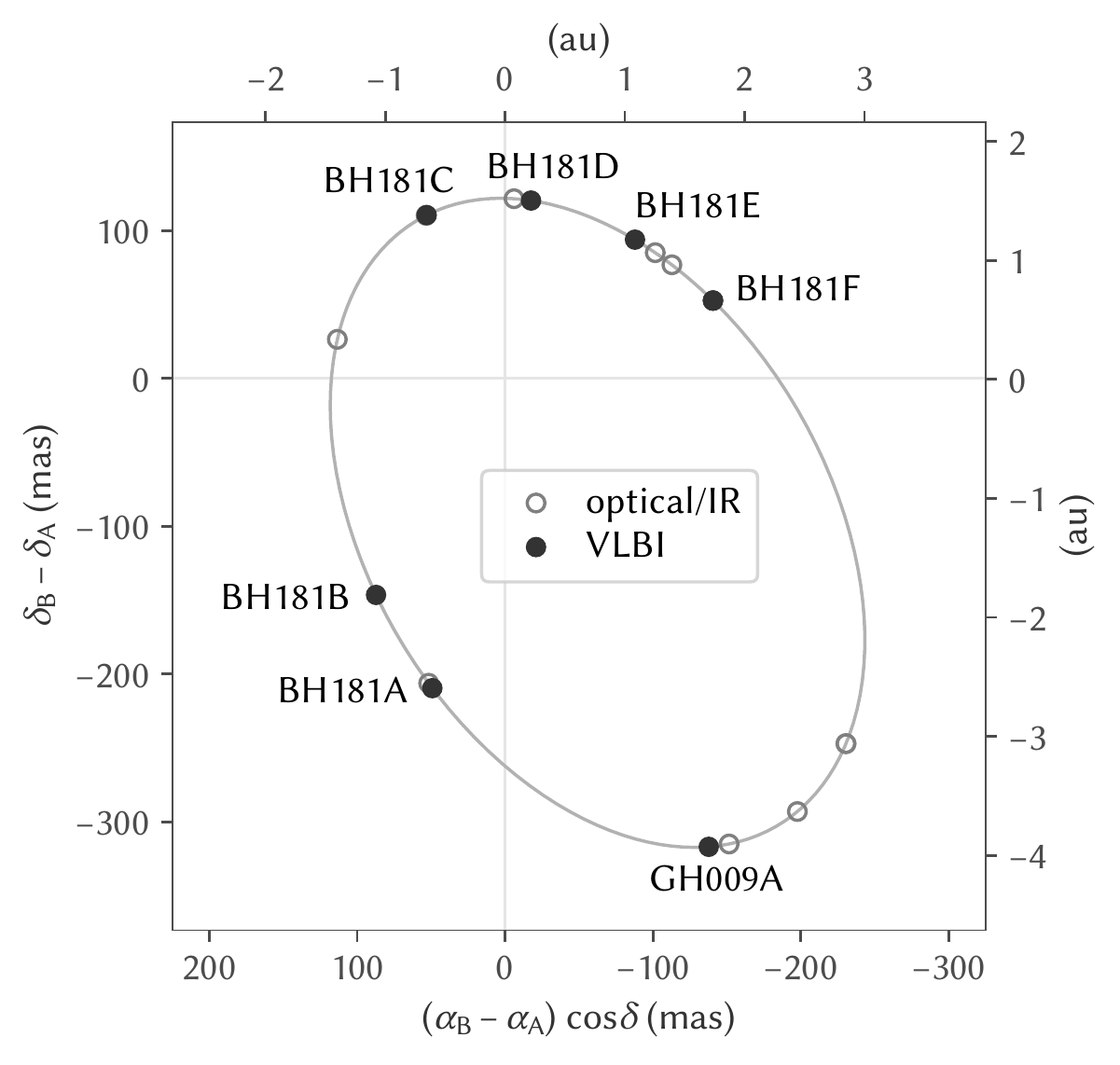}
\caption{Orbit of 2M J0746+2000B relative to A in R.A. ($(\alpha_{\text{B}}-\alpha_{\text{A}})\cos\delta$) and decl. ($\delta_{\text{B}}-\delta_{\text{A}}$), with the predicted positions at each of the seven VLBI epochs (labeled solid circles) and at the epochs of the previously published optical/IR observations in Table~\ref{tab:relastrom} (open circles). Note that circle sizes do not carry physical meaning; the VLBI measurements are too precise for their uncertainties to be illustrated at this scale.}
\label{fig:orbit_plot}
\end{figure}

\begin{deluxetable*}{@{\extracolsep{2mm}}lccccccc}
\rotate 
\tablecaption{Fitted parameters, their corresponding priors, and derived properties\label{tab:params}}

\tabletypesize{\scriptsize}
\tablecolumns{8}
\tablehead{\colhead{} & \multicolumn{3}{c}{Relative Optical/IR + Absolute VLBI} & \multicolumn{3}{c}{Absolute VLBI Only} & \colhead{}\\
\cline{2-4}\cline{5-7}
\colhead{Parameter} & \colhead{Median $\pm1\sigma$\tablenotemark{a}} & \colhead{$2\sigma$ Interval\tablenotemark{b}} & Best Fit & \colhead{Median $\pm1\sigma$\tablenotemark{a}} & \colhead{$2\sigma$ Interval\tablenotemark{b}} & Best Fit & \colhead{Prior}}
\startdata
\noalign{\vspace{0.8ex}}
\multicolumn{8}{c}{Fitted Parameters}\\
\noalign{\vspace{0.8ex}}
\hline
orbital period $P$ (yr) & $12.733\pm0.004$ & 12.726, 12.741 & 12.733 & $12.85_{-0.08}^{+0.09}$ & 12.66, 13.14 & 12.85 & $P^{-1}$\\
semi-major axis $a$ (mas) & $237.21_{-0.09}^{+0.08}$ & 237.04, 237.39 & 237.21 & $238.6_{-1.0}^{+1.1}$ & 236.1, 242.7 & 238.4 & $a^{-1}$\\
eccentricity $e$ & $0.48480\pm0.00029$ & 0.48422, 0.48535 & 0.48479 & $0.4874_{-0.0024}^{+0.0027}$ & 0.4818, 0.4956 & 0.4872 & 1, $0\leq e<1$\\
inclination $i$ ($\arcdeg$) & $138.36\pm0.06$ & 138.236, 138.480 & 138.36 & $138.19_{-0.22}^{+0.21}$ & 137.62, 138.63 & 138.24 & $\sin i$, $0\arcdeg<i<180\arcdeg$\\
pos. angle of asc. node $\Omega$ ($\arcdeg$) & $29.45\pm0.12$ & 29.21, 29.68 & 29.44 & $29.50\pm0.25$ & 29.00, 30.01 & 29.56 & 1\\
arg. of periastron $\omega$ ($\arcdeg$) & $355.94\pm0.16$ & 355.63, 356.26 & 355.93 & $355.95\pm0.24$ & 355.47, 356.42 & 355.89 & 1\\
periastron time $T_p$ (2015 Jul XX) & $24.2\pm0.5$ & 23.2, 25.3 & 24.2 & $23.4\pm1.0$ & 21.4, 25.4 & 22.9 & 1\\
A/B mass ratio $\xi\equiv M_A/M_B$ & $1.052\pm0.004$ & 1.044, 1.059 & 1.052 & $1.052\pm0.005$ & 1.042, 1.061 & 1.054 & $\xi^{-1}$\\
R.A. at J2010.0 $\alpha_0$ (07:46:XX)\tablenotemark{c} & $42.243483_{-0.000021}^{+0.000022}$ & 42.243441, 42.243526 & 42.243483 & $42.243451_{-0.000031}^{+0.000030}$ & 42.243389, 42.243510 & 42.243463 & 1\\
decl. at J2010.0 $\delta_0$ (+20:00:XX)\tablenotemark{c} & $31.4921\pm0.0005$ & 31.4911, 31.4932 & 31.4921 & $31.4923\pm0.0006$ & 31.4911, 31.4936 & 31.4926 & $\cos\delta_0$, $|\delta_0|\leq90\arcdeg$\\
R.A. proper motion $\dot{\alpha}\cos\delta$ (mas~yr$^{-1}$)\tablenotemark{c} & $-365.19_{-0.06}^{+0.05}$ & $-365.30$, $-365.09$ & $-365.20$ & $-365.12\pm0.08$ & $-365.27$, $-364.96$ & $-365.15$ & 1\\
decl. proper motion $\dot{\delta}$ (mas~yr$^{-1}$)\tablenotemark{c} & $-54.76_{-0.09}^{+0.10}$ & $-54.95$, $-54.57$ & $-54.76$ & $-54.81\pm0.11$ & $-55.04$, $-54.58$ & $-54.87$ & 1\\
parallax $\pi_{*}$ (mas) & $80.96_{-0.08}^{+0.09}$ & 80.80, 81.13 & 80.96 & $81.03\pm0.09$ & 80.85, 81.21 & 81.02 & $\pi_{*}^{-2}$\\
\hline
\noalign{\vspace{0.8ex}}
\multicolumn{8}{c}{Derived Properties}\\
\noalign{\vspace{0.8ex}}
\hline
distance $d$ (pc) & $12.352_{-0.013}^{+0.012}$ & 12.326, 12.377 & 12.352 & $12.341\pm0.014$ & 12.313, 12.368 & 12.342 & \nodata\\
semi-major axis $a$ (au) & $2.9300_{-0.0032}^{+0.0030}$ & 2.9237, 2.9364 & 2.9301 & $2.944_{-0.12}^{+0.14}$ & 2.915, 2.993 & 2.942 & \nodata\\
periastron distance $q$ (au) & $1.5095\pm0.0018$ & 1.5039, 1.5116 & 1.5096 & $1.509\pm0.005$ & 1.499, 1.519 & 1.508 & \nodata\\
apoastron distance $Q$ (au) & $4.351\pm0.005$ & 4.341, 4.360 & 4.351 & $4.379_{-0.23}^{+0.28}$ & 4.321, 4.476 & 4.375 & \nodata\\
total mass $M\equiv M_A+M_B$ ($M_\odot$) & $0.1552\pm0.0005$ & 0.1541, 0.1562 & 0.1552 & $0.1547\pm0.0013$ & 0.1521, 0.1573 & 0.1543 & \nodata\\
primary (A) mass $M_A$ ($M_\odot$) & $0.07954\pm0.00034$ & 0.07885, 0.08022 & 0.07955 & $0.0793\pm0.0007$ & 0.0780, 0.0806 & 0.0792 & \nodata\\
secondary (B) mass $M_B$ ($M_\odot$) & $0.07561\pm0.00025$ & 0.07512, 0.07611 & 0.07561 & $0.0754\pm0.0007$ & 0.0740, 0.0768 & 0.0751 & \nodata\\
\enddata
\tablenotetext{a}{$1\sigma$ uncertainties from the 15.866 and 84.134 percentiles of MCMC samples}
\tablenotetext{b}{2.275 and 97.725 percentiles of MCMC samples}
\tablenotetext{c}{ICRS reference position and motion of the system barycenter}
\end{deluxetable*}

\subsection{Model Fitting}

We constructed a two-body model of 2M J0746+2000A and B as gravitational point sources, with the system barycenter moving linearly with respect to the solar system. The model takes the six Keplerian orbital elements, the component mass ratio (A/B), as well as the position and proper motion of the system barycenter, and generates predictions for the position of B relative to A at each of the relative astrometry epochs in Table~\ref{tab:relastrom}, and for the absolute positions of A and B at each of the seven VLBI epochs. The relative astrometry observations used are identical to those used by \citet{dupuy2017} in their joint photocenter analysis, and similarly excludes one epoch (2003 February 18) from \citet{bouy2004} as an outlier, and a second epoch (2006 November 27) from \citet{konopacky2010} for which the imagery was not publicly available for re-analysis. Orbital coverage by the optical/IR and VLBI observations is shown in Figure~\ref{fig:orbit_plot}.

The model predictions are then compared with the actual observed values at each epoch to produce a likelihood. We consider astrometric uncertainty as normally distributed for all epochs except GH009A, and independent between epochs, other than the correlation between the in-beam corrected observations discussed earlier. Separation and position angle in the relative astrometry observations are treated as independent, while R.A. and decl. in the VLBI astrometry are treated as correlated by the computed covariance values.

The GH009A observation is downweighted by modeling its astrometric uncertainty as a 2D function whose radial cross sections are 1D Cauchy functions, with a $1\sigma$ error ellipse set by the stated $1\sigma$ uncertainties and covariance. This function was selected to be similar to a 2D normal distribution, should this observation be consistent with the others, while mitigating the observation's impact as an outlier otherwise, since the Cauchy function has much heavier tails and thus penalizes solutions with outliers to a lesser degree than the normal distribution does.

We additionally note that calibration errors in the BH181E epoch appear to have redistributed part of the flux for each of the A and B sources into a secondary peak offset $\sim$5~mas east of the measured primary peak in a subset of bands, likely causing the reported flux in Table~\ref{tab:posfit} to be an underestimate. Unlike GH009A, BH181E is surrounded by other, better calibrated epochs which already tightly constrain the source positions at BH181E. These surrounding observations are consistent only with the sources being located at the primary peaks. We therefore treat the measured primary peak positions as the properly calibrated source positions.

We found that an initial fit of the VLBI astrometry showed it to be systematically offset by $\sim$1$''$ to the southwest from the near IR photocenter trajectory from \citet{dupuy2017}. As the VLBI astrometry agrees with the J2015.5 position from the \emph{Gaia} DR2 data archive \citep{gaia2018}, we believe the offset to be an error in the absolute astrometric calibration used by \citet{dupuy2017} whose analysis largely depended only on accurate astrometry relative to the background stars (T.~J. Dupuy, private communication). We did not include the photocenter astrometry in our fit, and the observed shift was only used to inform an initial guess of the best fit parameters.

\begin{figure}
\centering
\includegraphics[width=\linewidth]{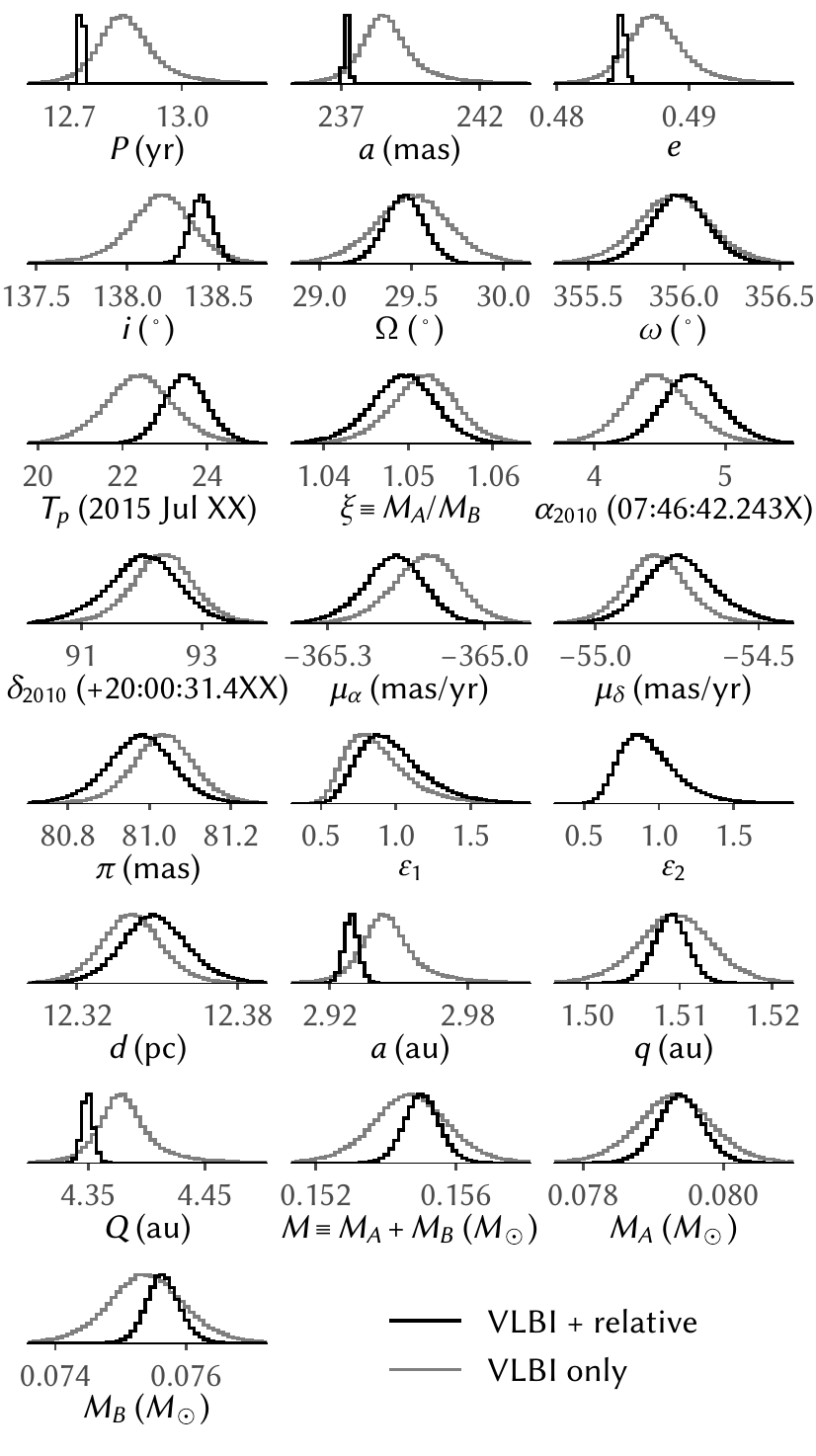}
\caption{Histograms of parameters in Table~\ref{tab:params} with MCMC samples from jointly fitting the relative optical/IR astrometry in Table~\ref{tab:relastrom} and the VLBI astrometry in Table~\ref{tab:posfit} (black), and from fitting the VLBI astrometry alone (gray). Corner plots showing the correlation between parameters in both sets of data are provided as the ancillary files \texttt{corner\_all.pdf} and \texttt{corner\_vlbi.pdf}.}
\label{fig:hist_compare}
\end{figure}

The \textsc{emcee} package \citep[version 2.2.1;][]{foreman-mackey2013} was then used to explore the parameter space through Markov-chain Monte Carlo sampling. First, the fit of \citet{dupuy2017} was used to locate a best fit solution maximizing the posterior function. One hundred walkers were then started with initial parameters concentrated in a tight ball surrounding this best fit with independent standard deviations one tenth the uncertainty stated by \citet{dupuy2017}. The walkers appeared to reach a steady state after 1000 steps, and both these first 1000 steps and the following 1000 steps were discarded to ensure the thoroughness of the burn-in. The next 100\,000 steps were then taken, with the parameters of every walker at every 100$^{\text{th}}$ step recorded as an MCMC sample, collectively representing the posterior distribution. Priors were selected to be isotropic in orientation (angles, restricted in range or wrapped as appropriate), uniform in time (periastron time) and volume (parallax), and log-uniform for other parameters bounded only to positive values. Characteristic statistics of this sample are presented for each of the fitted parameters in Table~\ref{tab:params} along with the corresponding priors. Statistics for several related properties derived from the fitted parameters are also computed from the sample and included in the table. Histograms showing all of these parameters and derived properties are presented in Figure~\ref{fig:hist_compare}.

With the extensive observation arc provided by the combined optical/IR and VLBI astrometry, all of the Keplerian orbital elements are tightly constrained to a relative precision of \result{$\sim$0.1\%} or better. The fitted mass ratio \result{$\xi\equiv M_A/M_B=1.052\pm0.004$} confirms that the primary A component is slightly more massive than the secondary B component. The total dynamical mass of 2M J0746+2000AB is calculated from these parameters to be \result{$0.1552\pm0.0005$~$M_\odot$}, split between the \result{$0.0795\pm0.0003$~$M_\odot$} primary and \result{$0.0756\pm0.0003$~$M_\odot$} secondary. These fitted values are consistent with the $1\sigma$ bounds for the same properties determined by \citet{dupuy2017} with near IR photocenter astrometry.

%
%
%

\subsection{Residuals Analysis}

\begin{figure*}
\centering
\includegraphics[width=0.85\linewidth]{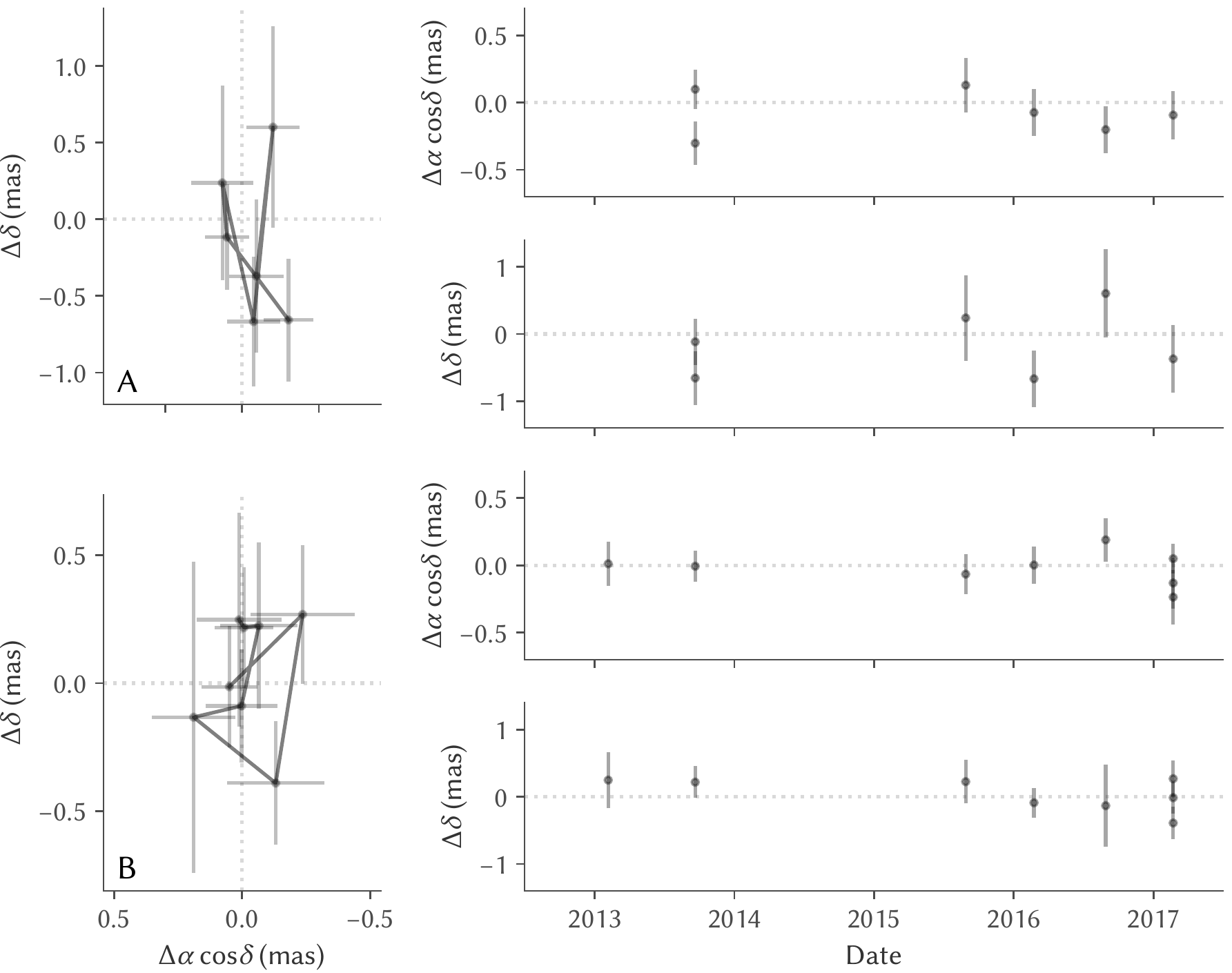}
\caption{VLBI astrometric residuals in R.A. ($\Delta\alpha\cos\delta$) and decl. ($\Delta\delta$) of the 2M J0746+2000A (upper half) and B (lower half) relative to the positions predicted by the best fit of the relative and VLBI astrometry. Gray lines connect the residuals of sequentially ordered observations in the 2D residual plots (left). The first epoch, GH009A, is excluded from this figure as an outlier with residuals of \result{$+1.1$~mas in R.A. and $-0.5$~mas in decl.}}
\label{fig:resid}
\end{figure*}

Astrometric residuals to the fitted two-body model can indicate or constrain the presence of additional unseen planets gravitationally perturbing the system. \citet{forbrich2016} used the absence of systematic residuals in the corresponding LSPM 1314+1320AB data to constrain the size and orbit of planets in that system. Here, we present a similar analysis for 2M J0746+2000AB.

Figure~\ref{fig:resid} compares the VLBI astrometry of 2M J0746+2000AB against the positions predicted by the best fit solution at the corresponding epochs. The GH009A observation, which had been downweighted as a possible outlier, was confirmed to be one by its large residuals of \result{$+1.1$~mas in R.A. and $-0.5$~mas in decl.} The rms of the residuals of the remaining observations is \result{0.14~mas} in the R.A. direction and \result{0.36~mas} in the decl. direction, which is consistent with the \result{$0.16\pm0.03$~mas} and \result{$0.40_{-0.09}^{+0.10}$~mas} expected from the formal astrometric uncertainties. For further validation, a separate orbital solution was computed with the additional free parameters $\varepsilon_1$ and $\varepsilon_2$ scaling the formal astrometric uncertainties of the VLBI and relative optical/IR astrometry, respectively. The fitted factors \result{$\varepsilon_1=1.1\pm0.2$} and \result{$\varepsilon_2=0.9\pm0.2$} are, again, consistent with the formal astrometric uncertainties.

The absence of clear and systematic deviations from the assumed two-body model constrains the mass and orbit size of planets orbiting either of the two components by the reflex motion a planet would impart on its parent. For a planet of mass $m_p$ orbiting a star of mass $M_*-m_p$ at a distance $a_p$, this reflex motion would have an amplitude $\delta r=m_p a_p/M_*$. The 2D rms of \result{0.27~mas} places a $3\sigma$ bound of \result{$\delta r<0.01$~au} on the reflex motion, corresponding to \result{$m_p a_p<0.9$~$M_{\text{jup}}$~au} for a planet orbiting either the A or B component. The presence of circumbinary planets cannot yet be effectively constrained, and require further VLBI observations to extend the observation arc beyond one orbital period.

\section{Conclusions}

VLBI observations of the L dwarf binary system 2M J0746+2000AB conducted at seven epochs over 2010--2017 reveal both components to be radio emitters, with the B component responsible for the bulk of the emission at 5~GHz frequency. Circularly polarized burst emission---a characteristic of the ECMI mechanism---was observed from both sources, with the 2.07~h periodic bursts identified by \citet{berger2009} confirmed to originate from the B component. A weak $\sim$3~h periodic signal was also detected from the A component, a value broadly consistent with the 3.3~h optical periodicity measured by \citet{harding2013b}.

This result marks the first instance of a UCD system observed with multiple radio emitting components. 2M J0746+2000AB is, moreover, only the second multiple UCD system probed with VLBI after the M7 binary LSPM J1314+1320AB, in which \citet{dupuy2016} observed only a single radio emitter. With $<$10\% of UCDs being appreciable radio sources, this finding hints at the possibility of a positive correlation in the presence of radio emission between the components of a multiple UCD system. VLBI investigation of additional radio emitting UCD systems will be necessary to robustly constrain any such correlation.

Additionally, VLBI astrometry anchors the system's relative orbit to an inertial reference frame, enabling a precise measurement of the mass ratio of the two components, and thus, their individual masses. Jointly fitting the VLBI astrometry with previously published relative optical/IR astrometry gives masses \result{$0.0795\pm0.0003$~$M_\odot$} and \result{$0.0756\pm0.0003$~$M_\odot$} for the A and B components, respectively, suggesting that both components likely exceed the minimum stellar mass threshold. These measurements represent the most precise individual mass estimates of UCDs to date, which follows from the high spatial resolution of VLBI imagery together with a combined observation arc extending nearly two decades---well over an orbital period. Residuals of the best fit orbital solution are broadly consistent with formal astrometric uncertainties, placing a $3\sigma$ bound of \result{$m_p a_p<0.9$~$M_{\text{jup}}$~au} on the mass and semi-major axis of planets orbiting either component.

\acknowledgments
{We thank M.~C. Liu for help isolating a systematic rotation in the existing relative astrometry, and both M.~C. Liu and T.~J. Dupuy, as well as an anonymous referee for reading this manuscript and providing valuable feedback. This material is based in part upon work supported by the National Science Foundation under Grant AST-1654815. The National Radio Astronomy Observatory is a facility of the National Science Foundation operated under cooperative agreement by Associated Universities, Inc. This work made use of the Swinburne University of Technology software correlator, developed as part of the Australian Major National Research Facilities Programme and operated under licence \citep{deller2011}. The Green Bank Observatory is a facility of the National Science Foundation operated under cooperative agreement by Associated Universities, Inc. The European VLBI Network is a joint facility of independent European, African, Asian, and North American radio astronomy institutes. Scientific results from data presented in this publication are derived from the following EVN project code(s): GH009.}

\facilities{EVN, GBT, VLBA}

\software{\textsc{aips} \citep{greisen1990}, \textsc{astropy} \citep{astropy2013}, \textsc{casa} \citep{jaeger2008}, \textsc{corner.py} \citep{foreman-mackey2016}, \textsc{emcee} \citep{foreman-mackey2013}, \textsc{matplotlib} \citep{hunter2007}, \textsc{python} \citep{vanrossum1995}}

\bibliography{ms}

\end{document}